# Angle-dependence of interlayer coupling in twisted transition metal dichalcogenide heterobilayers

W. T. Geng[a*], V. Wang[b], J. B. Lin[a], T. Ohno[a], J. Nara[a†]

[a] *National Institute for Materials Science, Tsukuba 305-0044, Japan.*
[b] *Department of Applied Physics, Xi'an University of Technology, Xi'an 710054, China*

We reveal by first-principles calculations that the interlayer binding in a twisted $MoS_2$/$MoTe_2$ heterobilayer decreases with increasing twist angle, due to the increase of the interlayer overlapping degree, a geometric quantity describing well the interlayer steric effect. The binding energy is found to be a Gaussian-like function of twist angle. The resistance to rotation, an analogue to the interlayer sliding barrier, can also be defined accordingly. In sharp contrast to the case of $MoS_2$ homobilayer, here the energy band gap reduces with increasing twist angle. We find a remarkable interlayer charge transfer from $MoTe_2$ to $MoS_2$ which enlarges the band gap, but this charge transfer weakens with greater twisting and interlayer overlapping degree. Our discovery provides a solid basis in twistronics and practical instruction in band structure engineering of van der Waals heterostructures.

---

[*] geng.wentong@nims.go.jp
[†] nara.jun@nims.go.jp

The weak van der Waals (vdW) interactions binding layered materials together set us free from dangling chemical bonds at the interfaces of heterostructure in semiconductor electronics. The rotational freedom in the configuration of a heterobilayer, the simplest form of vdW heterostructures, has already proved to be an astonishingly rich source in generating new quantum phenomena such as Hofstadter's spectra,[1,2] moiré excitons,[3,4,5,6] and Wigner crystal states.[7,8] Twistronics,[9] as a result, has been drawing extremely intensive research interest recently. Aside from these fascinating novel properties induced by moiré pattern, some fundamental features of the interlayer coupling, especially the twist-angle-dependence of which, also present challenges to first-principles theoretical studies that usually employ periodic boundary conditions to describe commensurate superlattices. Unlike homobilayers, which have commensurate superstructure for at least a set of (infinitely many) special discrete twist-angles, heterobilayers are inherently incommensurate due to the interlayer lattice mismatch, thus small strains have to be tolerated to keep the computation manageable. [10,11,12]

We note that first-principles density functional theory (DFT) calculations on vdW transition-metal dichalcogenides (TMDCs) heterobilayers with large superstructures were often performed only for the local regions with the three-fold rotational-symmetry, on a 1:1 supercell with an average lattice constant[13,14,15]. Since strain has strong influence on the electronic structures of TMDC,[16] full DFT treatment for the superlattice as a whole and at the same time to keep the strain to be reasonably small, is highly desirable in order to obtain more precise details of the Moiré potential and its impact on the interlayer vdW interactions. Since strain is practically inevitable in modeling a heterostructure, we propose a strategy to separate the effect of lattice misfit and twisting on the interlayer coupling.

Taking the $MoS_2/MoTe_2$ heterobilayer as an example, the lattice misfit between a supercell of $MoS_2$ $\{m\vec{a}_1, n\vec{a}_2\}$ with a hexagonal primitive cell $\{\vec{a}_1, \vec{a}_2\}$ and a supercell of $MoTe_2$ $\{p\vec{b}_1, q\vec{b}_2\}$ with a hexagonal primitive cell $\{\vec{b}_1, \vec{b}_2\}$ is

$$\delta a_M = \frac{a_M(MoS_2)}{a_M(MoTe_2)} - 1 = \frac{a \times \sqrt{m^2 + n^2 - mn}}{b \times \sqrt{p^2 + q^2 - pq}} - 1 \qquad (1)$$

where a and b are the length of $\vec{a}_1(\vec{a}_2)$ and $\vec{b}_1(\vec{b}_2)$. And the twist angle of this Moiré cell is

$$\theta_M = \left|\cos^{-1}\frac{2m-n}{2\sqrt{m^2+n^2-mn}} - \cos^{-1}\frac{2p-q}{2\sqrt{p^2+q^2-pq}}\right|. \quad (2)$$

We set a tolerance of interlayer lattice strain as 0.3%, within which there are dozens of stacking patterns. We choose seven patterns, with the twist angle falls in the range of $0 < \theta < \pi/6$ and the lattice misfit in a narrow range between -0.27% and -0.20%, as listed in Table 1. We note that in a bilayer system, the lattice strain after 2D optimization is not well defined because there is no sound reference system for each monolayer.[12]

Table 1. The interlayer lattice misfit, $\delta a_M$, twist angle, $\theta_M$, total number of atoms, $N_{atom}$ in the supercells used to model MoS$_2$/MoTe$_2$ heterobilayers formed by stacking a MoS$_2$ monolayer onto a MoTe$_2$ monolayer (near-R stacking). Also listed is the calculated average interlayer distance $d$ (in Å). For explanation of $m$, $n$, $p$, and $q$, see text.

| Stacking pattern $\{m,n\}/\{p,q\}$ | {10,1}/{9,1} | {10,4}/{9,4} | {9,4}/{8,3} | {10,4}/{9,5} | {9,4}/{8,5} | {11,5}/{9,1} | {9,4}/{7,0} |
|---|---|---|---|---|---|---|---|
| $\sqrt{m^2+n^2-mn}$ /$\sqrt{p^2+q^2-pq}$ | $\sqrt{91}/\sqrt{73}$ | $\sqrt{76}/\sqrt{61}$ | $\sqrt{61}/7$ | $\sqrt{76}/\sqrt{61}$ | $\sqrt{61}/7$ | $\sqrt{91}/\sqrt{73}$ | $\sqrt{61}/7$ |
| $\delta a_M$ | -0.27% | -0.25% | -0.20% | -0.25% | -0.20% | -0.27% | -0.20% |
| $\theta_M$ (°) | 0.61° | 2.92° | 4.54° | 10.26° | 11.88° | 21.18° | 26.33° |
| $N_{atom}$ | 492 | 411 | 330 | 411 | 330 | 492 | 330 |
| $d$ | 6.74 | 6.76 | 6.79 | 6.82 | 6.86 | 6.87 | 6.86 |

In each supercell, the length in $c$ axis was set to 30 Å, yielding a vacuum region of around 20 Å, large enough to minimize the interactions (except for the statistic Coulomb type) between mirror bilayers. To evaluate the dipole-dipole interaction caused by interlayer charge transfer, we have doubled the supercell size and included both a MoS$_2$/MoTe$_2$ and a MoTe$_2$/MoS$_2$ bilayer, separated by about 20 Å.

We performed the first-principles density functional theory (DFT) calculations using Vienna Ab initio Simulation Package.[17] The electron-ion interaction was described using projector augmented wave (PAW) method.[18] The exchange correlation between electrons was treated both with generalized gradient approximation (GGA) in the Perdew-Burke-Ernzerhof (PBE) form.[19] The non-bonding vdW interaction was incorporated by employing a semi-empirical correction scheme of Grimme's DFT-D3 method.[20] We used an energy cutoff of 400 eV for the plane wave basis set for all systems, both monolayers and bilayers, to ensure equal footing. The Brillouin-zone integration was performed within Monkhorst-Pack scheme[21] using *k* meshes of (2×2×1). The length of the supercell was fixed in the direction perpendicular to the bilayers, and was optimized in the other two. The structural relaxation is continued until the changes of the total energy of the supercell and forces on all the atoms are converged to less than $10^{-4}$ eV/cell and $2\times10^{-2}$ eV Å$^{-1}$ respectively. Due to the heavy weight of computation, spin-orbit coupling was not taken into account. We have generated the preprocessing initial atomic structure and the post-processing band structure of supercells using the VASPKIT code.[22]

We display in Fig.1 the optimized near-R stacking atomic structure of a $MoS_2$/$MoTe_2$ bilayer with a twist-angle of 4.54°. The configurations of other twist-angles can be found in the Supporting Information (Fig. S1.). We can find that, similar to the twist-free $MoTe_2$(9×9)/$MoS_2$(10×10) stacking we have studied,[12] the $MoS_2$ layer (upper) experiences more significant corrugation than the $MoTe_2$ layer (down), but to a lesser extent. As the twist-angle goes up from 0.61° to 26.33°, the $MoS_2$ layer becomes more and more flat and the interlayer distance increases from 6.74 Å to 6.86 Å (see Table 1).

**Fig. 1. Top (a) and side (b) views of a $MoS_2$/$MoTe_2$ bilayer with a twist-angle of 4.54° (near-R stacking). Small (coral), mediate (blue), and large (aquamarine) circles represent respectively S, Mo, and Te atoms.**

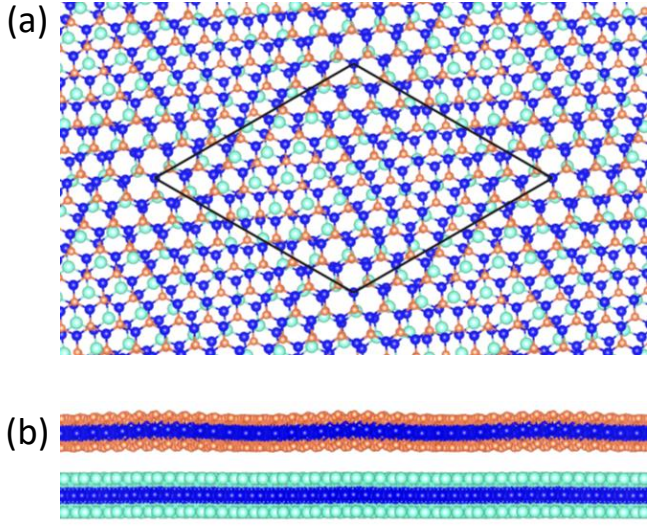

(a)

(b)

In Fig. 2, we show the calculated interlayer binding strength (dark cyan squares) for the heterobilayer with various twist-angle. The dipole-dipole interaction between bilayers induced by the charge transfer from MoS$_2$ to MoTe$_2$ has been cut out by employing two bilayers in one supercell with reflection symmetry. Without this correction, the binding strength will be 5.54-5.62 meV/Å$^2$ higher. It is found that the interlayer binding strength decreases with increasing twist-angles, and can be fairly well fitted by a Gaussian-type function with an assumed periodicity of π/3.

$$E_b(\theta) = E_b^0 + a \times e^{\frac{-sin^2[3(\theta-b)]}{2c^2}} = E_b^0 + a \times e^{\frac{cos(6\theta-2b)-1}{c^2}}, \qquad (3)$$

where $\theta$ is in unit of arc degree, $E_b^0$ =18.4 meV/Å$^2$, $a$=1.0 meV/Å$^2$, $b$=0°, and $c$=0.5 for a MoS$_2$/MoTe$_2$ heterobilayer. And the resistance to rotation, can be defined as

$$\tau(\theta) \equiv \frac{\partial E_b(\theta)}{\partial \theta} = -6a \sin\frac{6\theta-2b}{c^2}. \qquad (4)$$

It is easy to understand the correlation of larger interlayer distance and weaker interlayer binding from the nature of vdW interaction. However, it is not straightforward to understand why larger twist-angle yields weaker interlayer binding. The finding that the interlayer distance increases along with the twist-angle prompts us to assume the steric effect resulting from repulsive forces between overlapping electron clouds could be effectively described by a geometric quantity.

**Fig. 2.** The calculated interlayer binding strength, $E_b$, and the interlayer overlapping degree, $\Gamma$, in twisted $MoS_2/MoTe_2$ bilayers. The dotted lines are used to guide the eye, and the solid line is a Gaussian-type fitting of $E_b$.

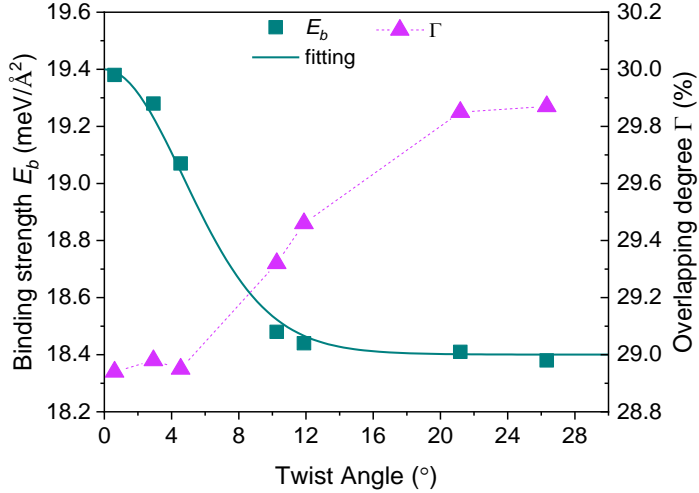

We noticed that viewed on-top of a bilayer (Fig. 1a), the overlap of atomic spheres in distinct layers varies significantly over the Moiré cell. This is always the case if there is lattice mismatch in a heterobilayer or a twisted homobilayer. Twist-free homobilayers are only extreme cases in which the degree of overlapping is homogeneous in unit of primitive cell. For instance, a graphene bilayer in the AA or AB stacking represent two typical cases. The overlapping is 100% in AA and 50% in AB, if we define the radius of a carbon atom being one half of the C-C bond length. Inspired by this observation, we here introduce a concept of *interlayer overlapping degree* (IOD) into a bilayer in general stacking patterns, taking the $MoS_2/MoTe_2$ heterobilayer as an example. The two atomic layers making the contact in this bilayer are S and Te and their overlapping is the interlayer overlapping. In each primitive (unit) cell in the upper S and lower Te layers, there is only one S or Te atom. Since there are no S-S and Te-Te bonds, the atomic size of S and Te cannot be well defined. For simplicity, we define the atomic radius of S and Te as $(\sqrt{3}/6)a$ of their respective primitive cell, similar to that of the C atoms in graphene. Then, the IOD of the heterobilayer can be defined as the ratio of the total overlapping area of S and Te atoms to the total area covered by S or Te in their respective atomic layers in the Moiré cell. We note that by definition, one S or Te atom covers $\sqrt{3}\pi/18$ of their respective primitive cell.

The calculated IOD for the optimized heterobilayer with different twist angles is displayed in Fig. 2 (purple triangles). Interestingly, we find that the IOD increases with the increasing twist-angle (with only the 2.92° case being slightly deviated), just opposite to the binding strength. This means that the IOD can serve well as a geometric descriptor of the repulsive steric effect, the variation of which influences the interlayer binding upon twisting.

**Fig. 3. Band structure of a MoS$_2$/MoTe$_2$ bilayer with a twist angle of 4.54° (a) and 26.33° (b).**

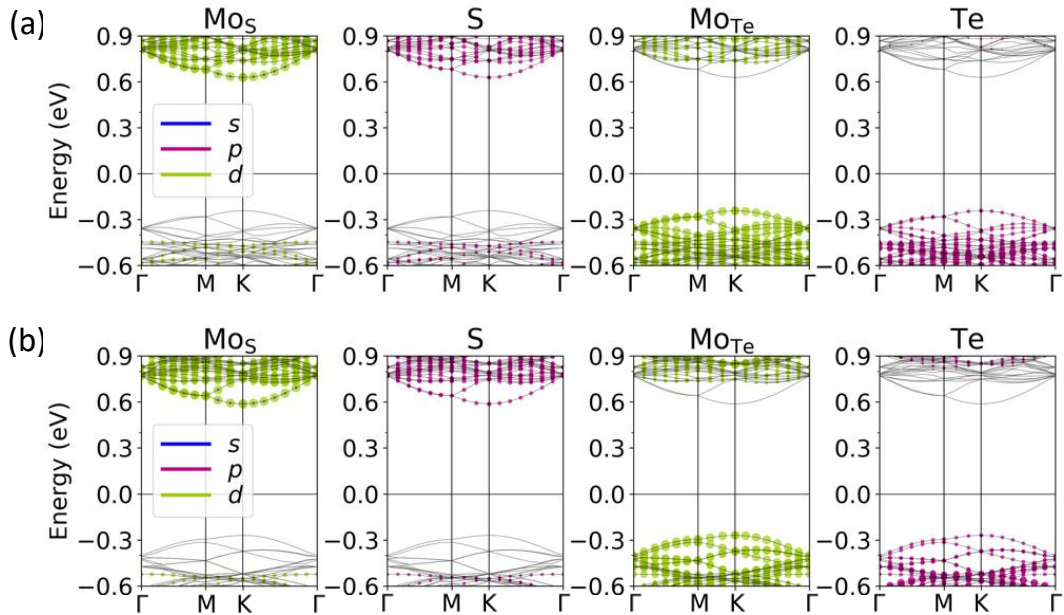

The calculated band structures near the Fermi level of a MoS$_2$/MoTe$_2$ heterobilayer with twist angle of 4.54° (panel a) and 26.33° (pane b) are plotted in Fig. 3. These two patterns have the same size of supercell, i.e., ($\sqrt{61} \times \sqrt{61}$)/(7 × 7) and the same interlayer lattice misfit (-0.20%), thus are well comparable. Projections are made on $s$, $p$, $d$ orbitals and also on each species. For MoTe$_2$(7 × 7), The valence band maximum (VBM) appears at the K point, which corresponds to the K point in the unfolded Brillouin zone of a free-standing MoTe$_2$ monolayer, and for MoS$_2$ ($\sqrt{61} \times \sqrt{61}$), the conduction band minimum (CBM) is located at the K point, corresponding also to the K point of a free-standing MoS$_2$ monolayer.[16] It is found that twist influences only slightly the band structure, due to the weakness of vdW interactions. The band structures of the heterobilayer with other twist-angles are also quite similar and therefore are not shown. From 4.54° to 26.33°, the VBM and CBM shifts down a little bit, and the band gap increases slightly from 0.857 to 0.843 eV. The valence band is mainly contributed by 4$d$ orbitals of Mo in the MoTe$_2$ layer (denoted by Mo$_{Te}$) and partially by 5$p$ states of Te. The conduction band consists of mainly 4$d$ of Mo in MoS$_2$ layer (denoted by Mo$_S$) and partially 3$p$ states of S. We note that in the twist-free $(10 \times 10)/(9 \times 9)$ stacking where the interlayer lattice misfit is +0.22%, the contribution of S-3$p$ is much less significant. [12] This is a strong indication that the effect of lattice strain on the bilayer electronic structure is more significant than does twist.

**Fig. 4. Band gap and interlayer charge transfer in twisted MoS$_2$/MoTe$_2$ bilayers. The solid line is a Gaussian-type fitting of $E_b$.**

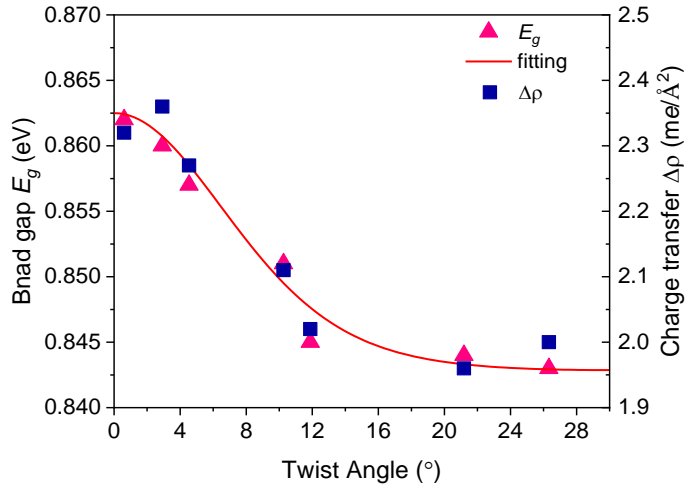

In Fig. 4, we display the calculated band gap of twisted MoS$_2$/MoTe$_2$ bilayers as a function of the twist angle (pink triangles). Our DFT calculations demonstrate unambiguously that the band gap decreases with increasing twist-angle. Similar to the binding energy, it can also be roughly fitted with a Gaussian-type function

$$E_g = E_g^0 + a \times e^{\frac{\cos(6\theta - 2b) - 1}{c^2}} = 0.843 + 0.02 \times e^{\frac{\cos(6\theta) - 1}{0.5}} \qquad (5)$$

This is in sharp contrast to the case of MoS$_2$ homobilayer[23,24] or MoS$_2$/WS$_2$ heterobilayer,[25] where the band gap increases with increasing twist-angle. The most significant difference in the interlayer electronic interaction between MoS$_2$/MoTe$_2$ and MoS$_2$/MoS$_2$ or MoS$_2$/WS$_2$ is that there is discernible interlayer charge transfer in the former case where the chalcogen element is different for two layers (see Fig.S2). And the interlayer charge transfer will induce electrostatic interlayer interaction, which in turn will give rise to energy level shift in opposite directions for the two constituent monolayers in a type II heterojunction. Using Bader analysis,[26] we have evaluated the interlayer charge transfer from MoTe$_2$ to MoS$_2$ in the twisted MoS$_2$/MoTe$_2$ bilayers, which are also shown in Fig. 3 (navy squares). Strikingly, we find that the interlayer charge transfer changes in phase with the band gap over the variation of twist-angle, roughly in proportion to each other.

Charge transfer from MoTe$_2$ to MoS$_2$ upshifts the electrostatic potential of MoS2 layer and downshifts that of MoTe$_2$. As a consequence, the energy level of the conduction band, which is contributed by MoS$_2$, moves up and the valence band, on the other hand, is leveled down. The band gap, therefore, will be enlarged. With the increase of twist-angle, the charge transfer reduces due to the enhanced interlayer overlapping, and hence a lessened band gap enlargement. To have a simple quantitative estimate, we take the bilayer as a parallel plate capacitor. The potential difference is

$$\Delta V = \frac{\Delta \rho \times d}{\varepsilon \times \varepsilon_0} \quad (6)$$

where $\Delta\rho, d, \varepsilon, \varepsilon_0$ are the two-dimensional charge density in the plate (interlayer charge transfer), inter-plate (interlayer) distance, relative dielectric constant and the vacuum dielectric constant, respectively. The separation distance of such a parallel plate capacitor is rather difficult to define, because the thickness (boundary) of a monolayer is not a well-defined quantity. We here take the size of this gap as the average interlayer distance between Mo$_S$ and Mo$_{Te}$ atomic layers (Table 1). The vertical dielectric constant of MoS$_2$ and MoTe$_2$ monolayers was reported to be 3.92 and 16.10 respectively.[27] We approximate the effective dielectric constant of the bilayer as the average of those two monolayers, 8.0. Then, $\Delta V$ is about 0.35 volt for 0.61° and 0.31 volt for 26.33°, and its reduction over the twist-angle range we have studied is about 0.04 volt. This magnitude is about two times as large as the band gap decrease (Fig.4). We note that the reduction of interlayer binding with increasing twist-angle (also interlayer distance) will bring about a band gap enlargement due to the dispersion reduction of the energy band near the Fermi level, as is demonstrated in the $MoS_2$ homobilayer.[23,24] Presumably, the impact of interlayer charge transfer on the band gap overcompensates the steric effect upon twisting the heterobilayer, leading to phenomenon that band gap decreases with the increasing twist angle. It has to be stressed that since there are strains (-0.20% ~ -0.27%) in the supercells we have employed, the change in the band gap, rather than band gap itself, is more meaningful to be discussed.

In summary, we have carried out first-principles DFT calculations on twisted $MoS_2$/$MoTe_2$ heterobilayers with a series of twist-angles. We have defined a geometric quantity, interlayer overlapping degree, to describe the repulsive interlayer steric effect. It is found that interlayer overlapping degree increases with the increasing twist-angle, hence the decrease of the interlayer binding. This relationship is a consequence of the non-bonding feature of the vdW interaction and not dependent on specific material. Therefore, it can be expected to hold in general cases of vdW bilayers. Interestingly, we find that the interlayer binding energy can be described by a Gaussian-like function of twist angle quite well in the small-angle range. As a consequence, the resistance to twist, which is an analogue to the interlayer sliding barrier, can be introduced to characterize the stability of a bilayer configuration with respect to the twist angle. Different to $MoS_2$ homobilayer, we find the energy band gap reduces with increasing twist angle. The underlying driving force turns out to be the interlayer charge transfer from $MoTe_2$ to $MoS_2$, which is weaker for greater twist where the interlayer overlapping degree is higher. The interlayer charge transfer enlarges the band gap through electrostatic effect. With a reduced charge transfer at large twist-angle, the band gap diminishes. What we have learned in this study add new basics to twistronics and will be instructive in band structure engineering of van der Waals heterostructures.

**ACKNOWLEDGMENTS**

This work was supported by Innovative Science and Technology Initiative for Security Grant Number JPJ004596, ATLA, Japan. The calculations were performed on Numerical Materials Simulator of NIMS. W.T.G. thanks Professor Qiang Gu for helpful discussions.